\documentclass[sigconf]{acmart}

\usepackage{graphicx}
\usepackage{tabularx}
\usepackage{xspace}
\usepackage{array}
\usepackage{makecell}
\usepackage{listings}
\usepackage{subcaption}
\usepackage{caption}
\usepackage{color,soul}
\usepackage{stfloats}
\usepackage[colorinlistoftodos,prependcaption,textsize=tiny]{todonotes}
\usepackage{gensymb}
\usepackage{balance}
\newcommand{\rr}[1]{#1}

\AtBeginDocument{%
  }

\copyrightyear{2026}
\acmYear{2026}
\setcopyright{cc}
\setcctype{by}
\acmConference[CHI '26]{Proceedings of the 2026 CHI Conference on Human Factors in Computing Systems}{April 13--17, 2026}{Barcelona, Spain}
\acmBooktitle{Proceedings of the 2026 CHI Conference on Human Factors in Computing Systems (CHI '26), April 13--17, 2026, Barcelona, Spain}
\acmDOI{10.1145/3772318.3791699}
\acmISBN{979-8-4007-2278-3/2026/04}

\begin{document}

\title{TactDeform: Finger Pad Deformation Inspired Spatial Tactile Feedback for Virtual Geometry Exploration}

\author{Yihao Dong}
\orcid{0009-0009-0719-3670}
\authornote{Both authors contributed equally to this research.}
\affiliation{%
  \department{School of Computer Science}
  \institution{The University of Sydney}
  \city{Sydney}
  \state{NSW}
  \country{Australia}
}
\email{yihao.dong@sydney.edu.au}

\author{Praneeth Bimsara Perera}
\orcid{0000-0001-9390-8381}
\authornotemark[1]
\affiliation{%
  \department{School of Computer Science}
  \institution{The University of Sydney}
  \city{Sydney}
  \state{NSW}
  \country{Australia}
}
\email{ppat0685@uni.sydney.edu.au}

\author{Chin-Teng Lin}
\affiliation{%
  \department{Australian AI Institute, School of Computer Science}
  \institution{University of Technology Sydney}
  \city{Sydney}
  \state{NSW}
  \country{Australia}}
\email{chin-teng.lin@uts.edu.au}
\orcid{0000-0001-8371-8197}

\author{Craig T Jin}
\orcid{0000-0003-4636-753X}
\affiliation{%
  \department{School of Electrical and Computer Engineering}
  \institution{The University of Sydney}
  \city{Sydney}
  \state{NSW}
  \country{Australia}
}
\email{craig.jin@sydney.edu.au}

\author{Anusha Withana}
\orcid{0000-0001-6587-1278}
\affiliation{%
  \department{School of Computer Science}
  \institution{The University of Sydney}
  \city{Sydney}
  \state{NSW}
  \country{Australia}
}
\affiliation{%
  \department{Sydney Nano Institute}
  \institution{The University of Sydney}
  \city{Sydney}
  \state{NSW}
  \country{Australia}
}
\email{anusha.withana@sydney.edu.au}

\renewcommand{\shortauthors}{Dong et al.}

\begin{abstract}
Spatial tactile feedback can enhance the realism of geometry exploration in virtual reality applications. Current vibrotactile approaches often face challenges with the spatial and temporal resolution needed to render different 3D geometries. Inspired by the natural deformation of finger pads when exploring 3D objects and surfaces, we propose TactDeform, a parametric approach to render spatio-temporal tactile patterns using a finger-worn electro-tactile interface. The system dynamically renders electro-tactile patterns based on both interaction contexts (approaching, contact, and sliding) and geometric contexts (geometric features and textures), emulating deformations that occur during real-world touch exploration. Results from a user study \rr{(N=24)} show that the proposed approach enabled high texture discrimination and geometric feature identification compared to a baseline. Informed by results from a free 3D-geometry exploration phase, we provide insights that can inform future tactile interface designs.

\end{abstract}



\begin{teaserfigure}
  \includegraphics[width=\textwidth]{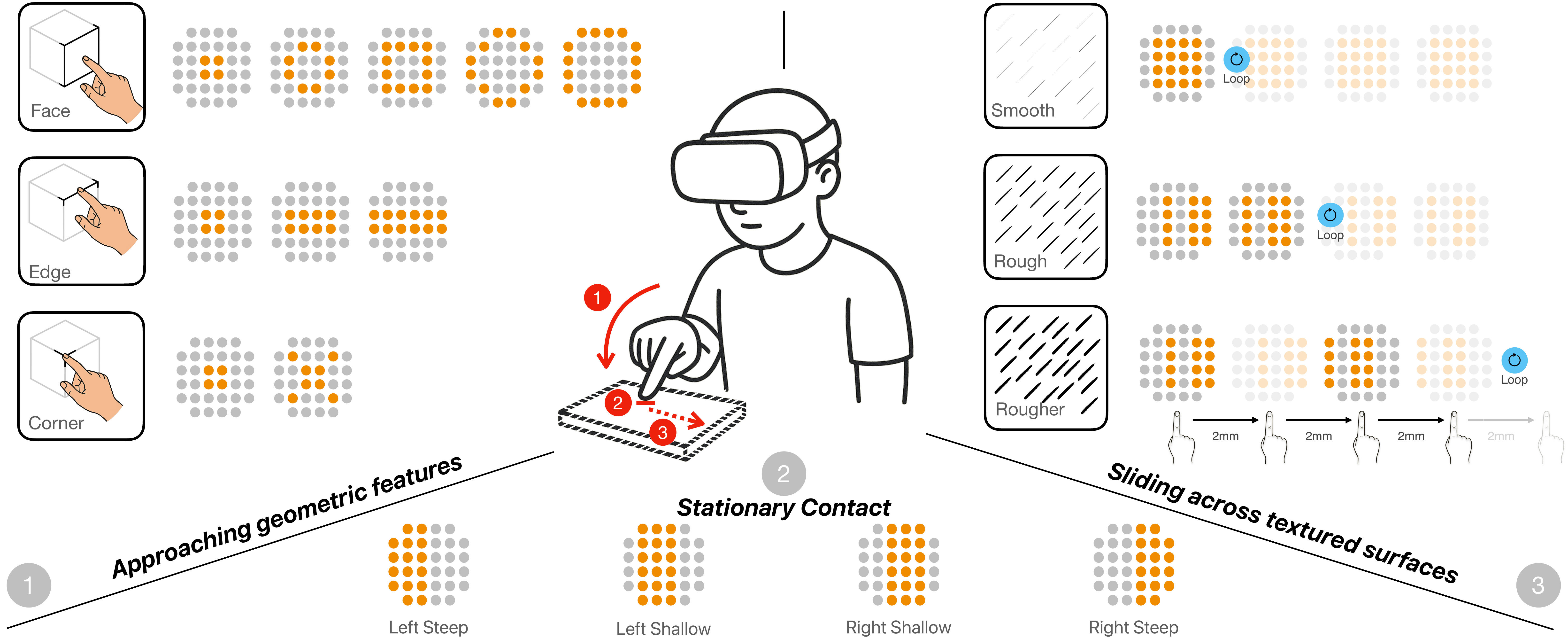}
  \caption{\textbf{TactDeform System Overview.} TactDeform renders spatio-temporal tactile patterns through a dual-context approach that adapts to both interaction contexts (approaching, contact, sliding) and geometric contexts (features and textures). The system emulates natural finger pad deformations using a 32-electrode array: (1) Approaching geometric features triggers expanding patterns for faces, directional patterns for edges, and concentrated patterns for corners; (2) Stationary contact shifts stimulation regions based on finger orientation; (3) Sliding across textured surfaces creates parametric pattern shifts synchronized with movement, smooth surfaces maintain static patterns, rough textures shift by 2-4mm per movement cycle.}
  \Description{The figure shows a central illustration of a person wearing a VR headset with their index finger positioned above a virtual object on a surface. Three interaction contexts are depicted around this central image:
  Top left (Approaching geometric features): Three rows showing Face, Edge, and Corner interactions. Each row displays a hand illustration touching a geometric shape alongside a 6 by 6 electrode grid pattern without the four corner nodes. Face patterns show orange electrodes expanding outward from the center. Edge patterns display orange electrodes in linear formations. Corner patterns concentrate orange activation at specific points.
  Bottom (Stationary Contact): Four electrode grid patterns labeled "Left Steep," "Left Shallow," "Right Shallow," and "Right Steep," showing how orange electrode activation shifts across the array based on finger orientation relative to the contact surface.
  Top right (Sliding across textured surfaces): Three texture levels—Smooth, Rough, and Rougher—each showing electrode patterns with blue loop arrows indicating rightward movement. Smooth shows static orange patterns, while Rough and Rougher display patterns that shift 2mm with each movement cycle, creating dynamic tactile feedback.
  Numbered arrows (1, 2, 3) connect these contexts, illustrating the natural progression of VR object exploration from approach to sustained contact to surface sliding.}
  \label{fig:teaser}
\end{teaserfigure}


\maketitle

\section{Introduction}\label{sec:intro}
Virtual reality (VR) has made significant strides toward replicating real-world sensory experiences.
This has established VR as a compelling platform for applications such as medical training~\cite{chelladurai_soundhapticvr_2024,shrestha_virtual_2025,zhou_coplayingvr_2024, Zhou2026OneBodyTwoMinds}, simulation~\cite{zhou_juggling_2025,cibulskis_tendon_2025,tsimbalistaia_-body_2025,katins_ad-blocked_2025,li_estimating_2025}, and interactive entertainment~\cite{yun_real-time_2025,ooms_haptic_2025,lee_skinhaptics_2025}.
% However, compared to the advancements in visual and auditory fidelity, tactile feedback remains a critical bottleneck for achieving realistic exploration and manipulation of virtual objects.
% interacting with geometry includes cutaneous (tactile) and kinesthetic (force) sensations.
\rr{However, compared to the fidelity in visual and auditory experiences, achieving realistic haptic experiences for rendering virtual geometry during exploration and manipulation remains a challenge.
This is due to the natural geometry interaction involving both high-resolution cutaneous (tactile)~\cite{kajimoto2004electro,kajimoto_electro-tactile_2016} and kinesthetic (force)~\cite{schorr_fingertip_2017,teng_pupop_2018,wang2019toward} sensations, which are difficult to replicate convincingly in VR, using a single haptic modality. To overcome this,~\citet{suga_presentation_2024} has explored multi-modal haptic interfaces that combine electro-tactile and kinesthetic feedback to enhance shape recognition in VR environments.}

\rr{However, due to physical constraints of the grounded force-feedback mechanism, this approach can only render small ``finger-sized'' virtual shapes, limiting its practicality for free-form geometry exploration in virtual environments. This is a natural limitation of kinaesthetic feedback devices such as exoskeletons~\cite{wang2019toward,teng_pupop_2018} or grounded force-feedback arms~\cite{endo2011five}, which tend to be heavy and physically restrictive. These factors significantly challenge the realistic and practical haptic rendering of 3D virtual geometries.
}

To address this, we propose \textit{TactDeform}, a parametric \rr{approach to render 3D virtual geometry using electro-tactile feedback. Electro-tactile interfaces forgo the need for bulky mechanical actuators, enabling thin and flexible form factors, and can deliver fast and localised tactile responses~\cite{withana2018tacttoo,yem2017wearable}.
\textit{TactDeform} is motivated by a specific insight: when we touch 3D geometries in the real world, our finger pad deforms in characteristic, feature-dependent ways. 
\textit{TactDeform} adapts electro-tactile stimulation patterns based} on the understanding of two interaction contexts. \rr{1) How users approach, make contact (touch), or explore a surface (e.g., slide across). And 2) what geometries they interact with; face, edge, corner, and textures (see Figure~\ref{fig:teaser}).} 
% For example, when approaching an edge, only part of the finger pad makes initial contact, creating a concentrated pressure region; when pressing a flat surface, the pad spreads out evenly; when sliding over a fine texture, dynamic micro-deformations occur across the skin.
% These deformation patterns are crucial tactile cues they engage different mechanoreceptors (e.g., SA1 receptors for sustained pressure vs. RA receptors for vibrations) that let us discern shapes and textures.
% \textit{TactDeform} introduces a parametric, electro-tactile based approach to emulate these deformations by delivering high-resolution, context-specific tactile feedback in VR.
\rr{Depending on these two contexts, \textit{TactDeform} emulates finger pad deformation using high-resolution, context-specific electro-tactile stimulation} using a lightweight finger-worn interface. \rr{Rather than physically deforming the skin with mechanical force feedback, we parametrically encode the spatial and temporal patterns of fingertip deformation that would naturally occur during real-world interactions.}
% allowing tactile feedback to adapt dynamically to both geometric features and interaction dynamics when interacting with objects in VR.}
% TactDeform depends on the understanding of two-contexts: the system uses (1) the interaction context (approaching, contact, or sliding) and (2) the geometric context (feature and texture). By parametrising these contexts and mapping them to specific spatial and temporal deformation patterns, TactDeform delivers localised electrical stimulation directly to the user's finger pad. This approach enables the system to generate a rich variety of tactile sensations as users actively explore 3D virtual geometries.

% By parametrising these contexts and mapping them to specific spatial and temporal deformation patterns, TactDeform delivers localised electrical stimulation directly to the user’s finger pad. 
% This approach enables the system to generate a rich variety of tactile sensations as users actively explore 3D virtual geometries, \rr{with only electro-tactile feedback}.
% This parametric approach emulates the natural deformation signatures that occur during real-world exploration, enabling users to distinguish edges from surfaces, perceive corners, and feel textures, all through a lightweight finger-worn interface.

We evaluated \textit{TactDeform}'s effectiveness in a user study involving \rr{24} participants across two phases. In the first phase, we assessed users' abilities to identify geometric features and textures using our parametric patterns. The second phase was a \rr{freeform} 3D geometry exploration study, spanning objects from simple primitives to complex forms, and comparing TactDeform to other tactile rendering approaches. This study provided qualitative insights into how users naturally integrate spatio-temporal tactile feedback into object exploration strategies. Our results demonstrate that TactDeform enables \rr{high accuracy in geometric feature identification (85.7\%) and texture discrimination (95.8\%),} and qualitative analysis reveals emergent, tactile-guided \rr{3D geometry} exploration behaviours.
In summary, this paper makes the following contributions:
\begin{itemize}
\item \textit{TactDeform}: a parametric approach that renders \rr{tactile feedback for 3D object interaction using context-specific} spatio-temporal \rr{electro-tactile feedback} patterns inspired by natural finger pad deformations.
% \rr{TactDeform: a novel parametric electro-tactile feedback approach that renders context specific spatio-temporal tactile patterns inspired by natural finger pad deformations during 3D object interaction.}
% \item A dual-context adaptation strategy for tactile rendering, enabling dynamic adjustment of electro-tactile feedback based on both interaction (approaching, contact, sliding) and geometric (features, textures) contexts.
% \item Empirical validation of TactDeform, including quantitative evidence for improved texture and feature identification, qualitative analysis of user exploration behaviours, and release of an open-source implementation with design guidelines for VR applications.
\item An open-source implementation \rr{of \textit{TactDeform} along with} design guidelines for \rr{integrating electro-tactile feedback into 3D geometry exploration in} VR applications \footnote{https://github.com/aid-lab-org/TactDeform}.
\item \rr{Insights from our qualitative analysis showing user exploration behaviours driven by spatial tactile feedback, highlighting the potential of electro-tactile interfaces for enhancing VR object interaction.}
\end{itemize}
\section{Related Work}

Prior research has demonstrated the critical role of haptics in enhancing user experience when interacting with virtual geometries. In this section, we review existing literature relevant to our work in three key areas: (1) using electro-tactile feedback is a viable tactile feedback option for spatial information rendering, (2) haptic rendering of 3D geometry in VR, and (3) parametric and context-based haptic rendering approaches.

\subsection{\rr{Electro-tactile Feedback for Spatial Rendering}}\label{subsec:eTactileFeedback}
\rr{Electro-tactile stimulation elicits tactile sensations by delivering controlled electrical current pulses using skin-surface electrodes, directly activating cutaneous sensory nerves~\cite{kajimoto2004electro,kajimoto2004smarttouch}. By modulating parameters such as amplitude, frequency, and waveform, these systems can evoke sensations including pressure, vibration, and texture~\cite{yem2016comparative,kajimoto2004electro,germani2013electro}, as well as complex percepts like softness~\cite{jingu2024shaping,suga2023softness,takei2015presentation}, temperature~\cite{saito23coldness}, and taste~\cite{mukashev2023tacttongue}. Compared to vibrotactile and force feedback, electro-tactile interfaces enable thin, lightweight, flexible form factors~\cite{withana2018tacttoo,yem2017wearable} with fast response~\cite{kajimoto_electro-tactile_2016}. This versatility makes them ideal for hand and finger-based applications~\cite{withana2018tacttoo,yem2017wearable}.}

\rr{Specifically in the context of rendering spatially distributed information, higher spatial resolution is critical.
Unlike vibrotactile actuators limited by mechanical crosstalk or force feedback requiring bulky linkages, electro-tactile interfaces can achieve millimeter-scale precision by independently controlling dense electrode arrays~\cite{Sobue_Optimal_2025}. By activating specific electrode subsets, these systems create localized stimulation patterns that can represent fine spatial details, aligning with natural spatial acuity in the skin~\cite{Sobue_Optimal_2025}.
Recent implementations incorporate these capabilities into virtual reality~\cite{suga_3d_2023,pamungkas2016electro}, accessibility~\cite{basdogan_review_2020,Teng_Seeing_2025}, and navigation~\cite{duente_shock_2024} applications.}

\subsection{\rr{Haptic Rendering of 3D Geometry in VR}}\label{subsec:HapticGeoInVR}
\rr{Haptic feedback enhances spatial understanding in VR and enables applications in medical training, CAD modelling, and accessibility interfaces~\cite{chelladurai_soundhapticvr_2024, rosa_influence_2015,feick_voxelhap_2023}.}
\rr{However, a fundamental challenge lies in encoding geometric features into perceptually meaningful patterns. Current approaches; whether force feedback~\cite{suga_3d_2023,teng_pupop_2018}, vibrotactile~\cite{martinez_identifying_2016,bau_teslatouch_2010}, or electro-tactile stimulation~\cite{kim_tactile_2013,peruzzini_electro-tactile_2012}; struggle with a critical requirement: accurate edge and corner representation~\cite{Ware_Search_2014}. Without clear differentiation between faces, edges, and corners, users cannot build coherent mental models of virtual objects. Empirical findings demonstrate this encoding failure. For instance, Suga et al.~\cite{suga_presentation_2024} found that shape discrimination accuracy dropped from 50\% (force + electro-tactile) to 25\% (electro-tactile alone), which is at chance level, revealing current electro-tactile patterns fail to convey geometric information despite high spatial resolution. Xu et al.~\cite{xu_tactile_2011} achieved only 56\% recognition for simple 2D shapes, indicating fundamental encoding inadequacy rather than merely 3D complexity. Osgouei et al.~\cite{osgouei_improving_2017} confirmed this limitation: users cannot naturally interpret geometric tactile patterns without extensive visual training. Even after training, recognition remains effortful rather than intuitive, indicating users must learn arbitrary perceptual mappings instead of using existing tactile processing.}

\rr{Current electro-tactile systems for geometric rendering employ two primary strategies. Uniform activation approaches activate all electrodes when the finger enters an object boundary~\cite{Teng_Seeing_2025}, providing binary contact feedback. Contact area mapping approaches activate individual electrodes based on proximity to virtual surfaces~\cite{Jiang_Designing_2024}, creating localized stimulation patterns. These encoding strategies focus on contact location but neglect the distinct tactile cues produced by different geometric features, resulting in lower discrimination performance.
Furthermore, sensations elicited by electro-tactile (tingling, vibration) might differ from their mechanically induced natural counterparts, which exacerbates this problem. Despite this discrepancy, users have managed to adapt to artificial sensations intuitively when electro-tactile stimulation is informed by the natural interaction context~\cite{jingu_double-sided_2023,jingu2024shaping}. This adaptation suggests that encoding strategies grounded in natural tactile processing may overcome perceptual mismatches, enabling effective spatial rendering despite inherent differences in sensation.}

\subsection{\rr{Parametric and Context-Based Haptic Rendering}}\label{subsec:parametricRendering}
\rr{The evolution of haptic interfaces from static to dynamic feedback represents a fundamental shift in understanding how tactile information should be encoded. This is evident from recent advances in motion-coupled, force-sensitive, and context-aware systems demonstrating the importance of adaptive rendering~\cite{strohmeier_pulse_2018,jingu2024shaping,Lin_Slip_2025,sabnis_motion-coupled_2025,kim_pro-tact_2024}. Early haptic systems generated fixed patterns regardless of user interaction, an approach achieving sub-60\% recognition even after training~\cite{osgouei_improving_2017}. This limited performance stems from a mismatch between static patterns and the dynamic nature of tactile exploration, where sensations evolve based on movement velocity, contact pressure, and surface traversal.}

\rr{The recognition that haptic perception is inherently active has driven the development of motion-coupled rendering systems.~\citet{strohmeier_pulse_2018} demonstrated that identical vibrotactile pulses produce entirely different texture perceptions depending on synchronization with finger movement; smooth when aligned, rough when offset. This reveals temporal coupling as fundamental rather than merely enhancing. Recent visual-haptic work~\cite{sabnis_motion-coupled_2025,dong_just_2025} extends this principle, demonstrating that motion-coupled feedback enhances realism across modalities and that desynchronization alters sensations, thereby establishing temporal synchronization as a core principle for believable sensory simulation.}

\rr{Building beyond motion coupling, researchers have explored multi-parametric systems where patterns adapt to multiple interaction variables simultaneously~\cite{yu_irontex_2024, yu_fabricating_2024, Zhou2026SRLProxemics}.~\citet{jingu2024shaping} modulated the number of electro-tactile pulse grains based on contact force to create compliance illusions that respond to user pressure, while~\citet{Lin_Slip_2025} combine force sensing with slip simulation to convey weight. Similarly, recent wearable haptic systems~\cite{yao_encoding_2022} demonstrate that parametric rendering, systematic variation of stimulation based on interaction parameters, enables more realistic and intuitive perceptual experiences than fixed patterns.}

\rr{However, existing parametric approaches focus on single interaction contexts; motion for texture, force for compliance, without simultaneously considering geometric context. No current system adapts feedback based on both how users touch (interaction parameters) and what they touch (geometric features), limiting their ability to convey complex 3D shapes where interaction and geometry are fundamentally coupled. This directly motivates our adaptation strategy in TactDeform, enabling dynamic adjustment of electro-tactile feedback to emulate fingertip deformation, based on both interaction (approaching, contact, sliding) and geometric (features, textures) contexts to render 3D geometry in VR.
}

\section{TactDeform Concept}\label{sec:system}

The fundamental principle underlying TactDeform is that different interaction and geometric contexts produce distinct, reproducible deformation patterns on the finger pad. These deformation patterns arise from the mechanical properties of finger pad tissue and are detected mainly by four types of specialized mechanoreceptors that encode different aspects of tactile information~\cite{banes_mechanoreception_1995}.

\subsection{Encoding Virtual Geometry via Fingertip Deformation Emulating Electro-Tactile Feedback}
\rr{Mechanoreceptors jointly encode shape and texture through characteristic patterns of skin deformation rather than through explicit geometric variables~\cite{johansson_coding_2009}. Slowly adapting type I receptors (SA-I/Merkel cells) respond to sustained, localized deformation and fine spatial features such as edges and curvature via their dense distribution and small receptive fields~\cite{johansson_coding_2009}. Rapidly adapting receptors (RA-I/Meissner corpuscles) are instead tuned to dynamic changes in skin deformation during contact and sliding, providing temporal cues linked to motion and surface transitions~\cite{bensmaia_tactile_2008}. During texture exploration, perceived roughness emerges from the combined spatial pattern of indentation and the temporal pattern of vibration generated as the skin moves over the surface~\cite{weber_spatial_2013}.}

\begin{figure*}
    \centering
    \includegraphics[width=1\linewidth]{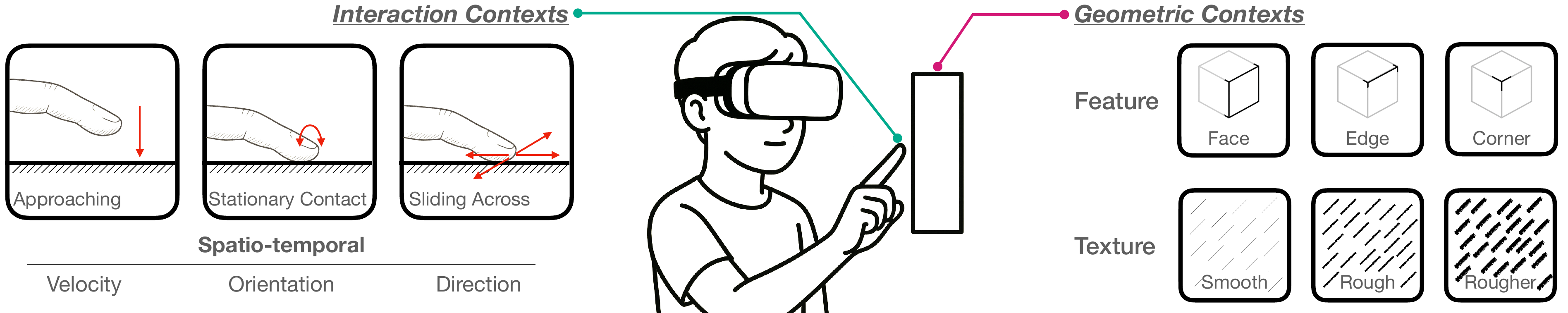}
    \caption{\textbf{TactDeform Concept.} TactDeform leverages a dual-context approach that combines interaction contexts (left) and geometric contexts (right) to generate parametric spatio-temporal tactile patterns. Interaction contexts include approaching, stationary contact, and sliding across surfaces, with associated parameters of velocity, orientation, and direction. Geometric contexts comprise feature types (faces, edges, corners) and texture levels (smooth, rough, rougher). This approach dynamically combines these contexts to render appropriate electro-tactile patterns that emulate natural finger pad deformations during virtual object exploration.}
    \Description{A conceptual diagram illustrating TactDeform. The centre shows a person wearing a VR headset and interacting with virtual objects using their index finger. On the left, three boxes demonstrate interaction contexts: Approaching (finger moving vertically toward an invisible object with downward arrow), Stationary Contact (finger touching surface with rotation indicator), and Sliding Across (finger moving horizontally across surface with directional arrows). Below these are labels for spatio-temporal parameters: Velocity, Orientation, and Direction. On the right, geometric contexts are divided into two categories: Feature types showing Face (cube with flat surface highlighted), Edge (cube with edge highlighted), and Corner (cube with corner highlighted); and Texture levels showing Smooth (no pattern), Rough (diagonal hatching), and Rougher (denser diagonal hatching). Cyan and magenta arrows connect the interaction and geometric contexts to the central user, indicating how both inform the tactile rendering.}
    \label{fig:concept}
\end{figure*}

\rr{Informed by this, TactDeform generates tactile patterns to emulate characteristics of a gradual increment of deformation (RA-I response) during the approach to a 3D object by dynamically expanding patterns,
orientation-dependent deformation during stationary contact (SA-I response) by localised static patterns, velocity and texture dependent deformation during sliding interactions (RA-I responses) by shifting or moving activation patterns as described in Section~\ref{sec:spatio_temporal_pattern_generation}.}
\rr{This deformation-based strategy contrasts with~\citet{suga_presentation_2024}, who map visual-geometric properties directly to intensity levels and require additional force feedback to reach higher recognition accuracy, effectively treating tactile stimulation as a symbolic channel.}
\rr{TactDeform achieves this through a dual-context approach that combines interaction contexts (approaching, contacting, or sliding across an object) with geometric contexts (features or textures) to generate parametric spatio-temporal patterns, as illustrated in Figure~\ref{fig:concept}.}

\subsection{Interaction Contexts:} The system recognizes three distinct interaction contexts that correspond to natural exploration behaviors (Figure~\ref{fig:concept}, left). \textit{Approaching:} When the finger pad initially approaches and makes contact with a virtual object, TactDeform renders spatio-temporal patterns that emulate the progressive deformation occurring during initial contact. These patterns convey the object's primary geometric features (faces, edges, corners) through expanding stimulation that mimics how pressure spreads across the finger pad during approach. \textit{Contact:} During sustained stationary contact, the system delivers orientation-specific patterns that emulate how the finger pad deforms based on the contact angle with the surface. These patterns maintain spatial awareness of the contact region while adapting to the finger's orientation relative to the object geometry. \textit{Sliding:} When the finger moves across virtual surfaces, TactDeform generates dynamic patterns that emulate the shear deformations and friction-induced variations experienced during surface exploration. These patterns convey texture information through parametric modulation of the stimulation based on surface roughness.

\subsection{Geometric Contexts:} Within each interaction context, TactDeform parameterizes geometric features and textures (Figure~\ref{fig:concept}, right).
\textit{Geometric Features:} The system classifies local geometry into three categories based on their deformation signatures:
\textit{Faces:} Planar or gently curved surfaces that create distributed deformation across the finger pad. 
\textit{Edges:} Linear boundaries that concentrate deformation along a specific direction. 
\textit{Corners:} Point features where multiple surfaces converge, creating localized deformation.
\textit{Surface Textures:} The system renders three parametric roughness levels that modulate the spatio-temporal patterns during sliding interactions, emulating how surface textures create varying friction and deformation dynamics.

\subsection{Parametric Pattern Design}
TactDeform's parametric approach translates finger pad deformation physics into spatio-temporal electro-tactile patterns through systematic mapping of deformation characteristics to stimulation parameters, as illustrated in Figure~\ref{fig:teaser}.

\subsubsection{Deformation-to-Stimulation Mapping:} 
\begin{figure*}
    \centering
    \includegraphics[width=1\linewidth]{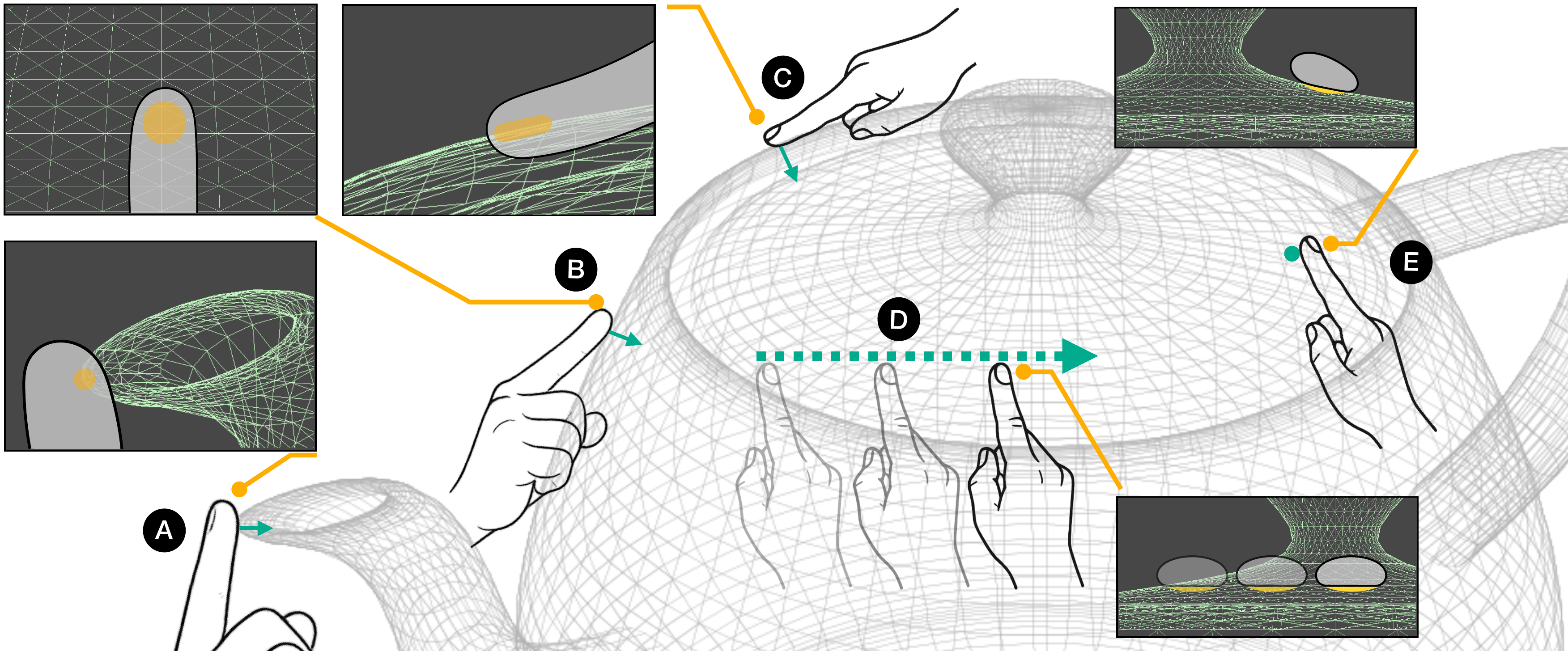}
    \caption{
    Finger pad deformation visualization during different interaction contexts. 
    (A) Finger pad approaches the tip of a teapot (a corner). 
    (B) Finger pad approaches the body of the teapot (a face). 
    (C) Finger pad approaches the ridge between the body and lid of the teapot (an edge). 
    (D) Moving across the lid of the teapot, showing progressive deformation patterns across the contact area. 
    (E) Finger pad remains stationary on the lid of the teapot, biased slightly to the left, with finger pad deformation reflecting this angular bias.
    }
    \label{fig:deformation}
    \Description{This figure illustrates finger pad deformation patterns during various tactile interactions with a 3D teapot model. The central image shows a large wireframe teapot rendered in gray mesh. Five labeled panels (A-E) surround the teapot, each showing close-up views of finger-object interactions with corresponding deformation visualizations. Panel A (top left) shows a finger approaching the teapot's spout tip, representing corner interaction. Panel B (middle left) depicts finger contact with the curved teapot body, representing face interaction. Panel C (top center) shows a finger approaching from above, illustrating edge interaction at the body-lid junction. Panel D (center) displays a sequence of four finger positions moving horizontally across the teapot surface, with a green dotted arrow indicating the movement direction, demonstrating sliding interaction patterns. Panel E (right) shows a finger positioned at an angle on the teapot surface, with blue and red directional arrows indicating the angular orientation bias. Each panel includes yellow connection lines linking to corresponding regions on the main teapot model. The visualization uses green wireframe overlays to represent the deformation fields and contact areas around each finger position.}
\end{figure*}

Each geometric feature produces a characteristic deformation signature (shown in Figure~\ref{fig:deformation}) that TactDeform parameterizes into corresponding electro-tactile patterns, shown in Figure~\ref{fig:teaser}.1 (top left):
\textit{Face Deformation:} When the finger pad contacts a planar surface, pressure distributes radially outward from the initial contact point. TactDeform emulates this through a diverging ring pattern that expands from the center of the electrode array, with parametric control over expansion rate.
\textit{Edge Deformation:} Linear features create directional pressure concentrations along the edge line. The system renders this as a parametric line pattern that extends from the center, with orientation parameters matching the edge direction parameters reflecting the edge \rr{shape}.
\textit{Corner Deformation:} Point contacts produce highly localized, intense deformation at the contact point. TactDeform renders this through concentrated stimulation at the array center, with parametric falloff controlling the spatial extent of the sensation.

\subsubsection{Spatio-Temporal Pattern Generation:}\label{sec:spatio_temporal_pattern_generation}

TactDeform generates dynamic electro-tactile patterns that adapt to both interaction velocity and geometric context, ensuring spatial correspondence between virtual interactions and tactile sensations. The system employs velocity-dependent parameterization across all three interaction contexts, with mathematical formulations that maintain perceptual consistency during exploration.

\textit{Approaching Context:} The system generates time-varying patterns where each frame represents one electrode spacing of deformation progression, matching the physical electrode array geometry (2mm center-to-center spacing in our implementation) shown in Figure~\ref{fig:teaser}.1 (top left). The expansion rate $r$ (in electrode spacings per second) is parametrically linked to approach velocity through:
\begin{equation}
r = \alpha \cdot \frac{v_{approach}}{d_{electrode}}
\rr{\label{eq:approach_rate}}
\end{equation}
where $v_{approach}$ represents the instantaneous finger pad velocity toward the virtual surface (mm/s), $d_{electrode} = 2.0$ mm denotes the center-to-center electrode spacing, and $\alpha$ is a scaling coefficient set to 1.0 \rr{through pilot testing (Section~\ref{subsec:pilotStudy})} for direct spatial correspondence. This formulation ensures that one electrode spacing of physical finger movement produces one electrode spacing of pattern expansion.

The temporal frame advancement interval $\Delta t$ between successive pattern frames is computed as:
\begin{equation}
\Delta t = \frac{d_{electrode}}{v_{approach}}
\rr{\label{eq:approach_interval}}
\end{equation}
This inverse relationship ensures faster approaches trigger more rapid pattern updates, while slower exploration produces gradual pattern evolution. In our Unity implementation, the pattern frame index $n(t)$ at time $t$ is determined through velocity integration:
\begin{equation}
n(t) = \lfloor \int_{0}^{t} \frac{v_{approach}(\tau)}{d_{electrode}} d\tau \rfloor
\rr{\label{eq:approach_frame}}
\end{equation}
The pattern selection (face, edge, or corner) is determined by the local geometry at the contact point, with each geometric feature maintaining its characteristic spatial pattern while sharing this unified velocity-dependent expansion mechanism.

\textit{Contact Context:} For stationary interactions, TactDeform parameterizes finger orientation based on the angle between a ray from the finger pad center to the contact point and the surface tangent at that point. After projecting onto the horizontal plane (flattening the z-axis), this angle $\theta$ is discretized into four orientation levels:
\begin{itemize}
\item \textit{Steep Left:} $\theta \in [-90\degree, -30\degree)$ - stimulation shifts to leftmost electrode region
\item \textit{Shallow Left:} $\theta \in [-30\degree, 0\degree)$ - stimulation shifts to left-center region  
\item \textit{Shallow Right:} $\theta \in [0\degree, 30\degree)$ - stimulation shifts to right-center region
\item \textit{Steep Right:} $\theta \in [30\degree, 90\degree]$ - stimulation shifts to rightmost electrode region
\end{itemize}

\noindent This parametric mapping, demonstrated in Figure~\ref{fig:teaser}.2 (bottom), is inspired by how different finger pad regions experience varying pressure distributions based on contact angle, as shown in Figure~\ref{fig:deformation}E. The orientation quantization into four discrete levels balances perceptual clarity with smooth transitions during exploration. We focus on left-to-right orientation mapping rather than top-to-bottom variations, as our electrode array covers the primary contact area of the fingertip while the top section remains uncovered. 

\noindent The orientation-specific patterns operate in combination with the underlying geometric feature patterns through spatial modulation:
\begin{equation}
P_{contact}(x,y) = P_{base}(x + \Delta x(\theta), y)
\rr{\label{eq:contact_pattern}}
\end{equation}
where $P_{base}$ represents the base geometric pattern (face, edge, or corner), and $\Delta x(\theta)$ is the orientation-dependent horizontal shift. This ensures that geometric information is preserved while the stimulation region adapts to finger orientation, maintaining both shape and contact angle information simultaneously.

\textit{Sliding Context:} Surface texture is parameterized through pattern shift mechanisms inspired by friction-induced deformations, illustrated in Figure~\ref{fig:teaser}.3 (top right) and Figure~\ref{fig:deformation}D. The pattern shift adapts to both the magnitude and direction of finger movement. \rr{Movement direction is quantized into 8 directions (forward, backward, left, right, plus four diagonal corners), established through pilot testing (Section~\ref{subsec:pilotStudy}) to balance directional sensitivity with perceptual stability.} The shift rate $s$ (in pattern shifts per second) follows the same velocity-dependent principle:
\begin{equation}
s = \beta \cdot \frac{v_{slide}}{d_{electrode} \cdot k_{texture}}
\rr{\label{eq:texture_rate}}
\end{equation}
where $v_{slide}$ represents the lateral finger velocity across the surface (mm/s), $k_{texture}$ is the texture-dependent shift multiplier, and $\beta$ \rr{is the directional coefficient ($\beta = +1$ for opposite-direction shifts, $\beta = -1$ for same-direction shifts). Through pilot testing (Section~\ref{subsec:pilotStudy}), we found opposite-direction shifts ($\beta = +1$) produced more realistic friction sensations.} The texture parameters are defined as:
\begin{itemize}
\item \textit{Smooth (Level 1):} $k_{texture} = 0$ - zero shift parameter, static stimulation regardless of movement
\item \textit{Rough (Level 2):} $k_{texture} = 1$ - one electrode spacing shift (2mm) synchronized with finger movement
\item \textit{Rougher (Level 3):} $k_{texture} = 2$ - two electrode spacing shift (4mm) with increased spatial period
\end{itemize}

\noindent The temporal shift interval for texture patterns becomes:
\begin{equation}
\Delta t_{shift} = \frac{d_{electrode} \cdot k_{texture}}{v_{slide}}
\rr{\label{eq:texture_interval}}
\end{equation}
For example, sliding at 20 mm/s across a rough surface generates pattern shifts at 10 Hz, while the same velocity on a rougher surface produces 5 Hz shifts, creating distinct texture sensations.

\noindent Importantly, these texture patterns operate in combination with the underlying geometric feature patterns. When sliding across a rough edge, the system multiplies the rough sliding pattern with the edge pattern, preserving both geometric and textural information through a unified parametric framework:
\begin{equation}
P(t) = P_{base} \cdot \lfloor \int_{0}^{t} \frac{v_{slide}(\tau)}{d_{electrode} \cdot k_{texture}} d\tau \rfloor
\rr{\label{eq:texture_pattern}}
\end{equation}
where $P_{base}$ represents the base geometric pattern. Based on movement direction, the non-smooth patterns exhibit directional coupling: rightward finger movement triggers leftward pattern shift, representing the relative motion between finger pad and surface.

The parametric mappings were iteratively refined through pilot testing \rr{(Section~\ref{subsec:pilotStudy}) to ensure that the spatio-temporal patterns effectively convey natural finger pad deformation characteristics while maintaining perceptual clarity and user comfort.} The final parameter sets were validated through the formal user evaluation described in Section~\ref{sec:userStudy}.
\begin{figure*}
    \centering
    \includegraphics[width=1\linewidth]{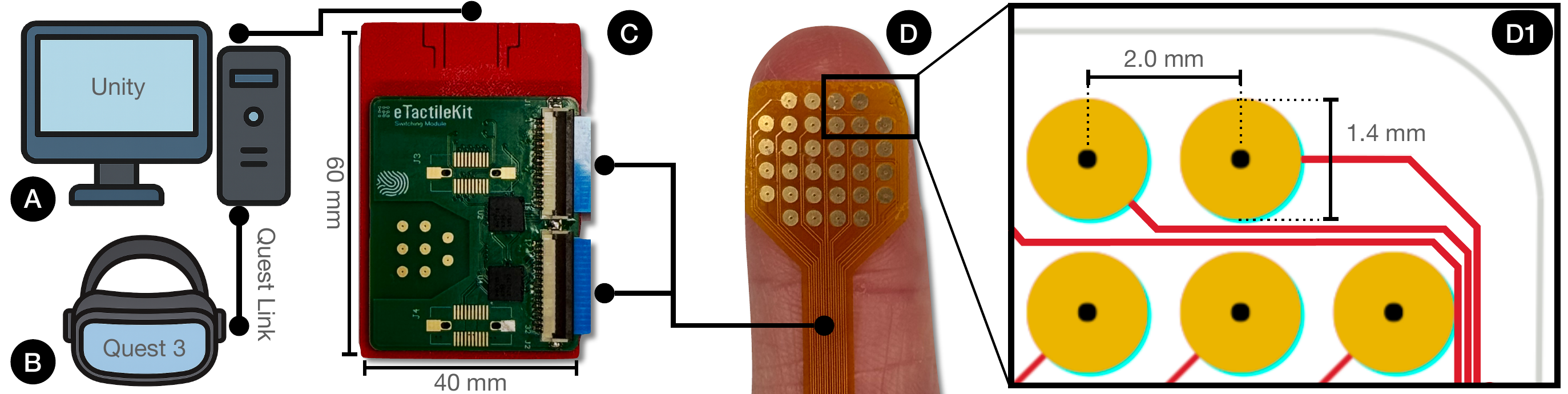}
    \caption{The system integrates four components: (A) Desktop PC running Unity application connected via Quest Link to (B) Meta Quest 3 HMD for hand tracking and visual rendering, (C) Control circuit in a 3D-printed enclosure communicating via serial connection, (D) Flexible PCB tactile interface attached to the index finger pad with 32 electrodes in a 6$\times$6 configuration, and (D1) Enlarged view showing electrode specifications with 1.4mm diameter and 2.0mm center-to-center spacing. The flexible PCB extends from the palm side for minimal movement interference. Medical tape for finger attachment not shown for clarity.}
    \Description{This figure illustrates the complete hardware architecture of the TactDeform system through a left-to-right component diagram. On the far left, a desktop computer monitor displays "Unity" with a tower PC next to it. The PC connects via Quest Link to a Meta Quest 3 head-mounted display shown below the PC. A serial connection links the PC to the central component: an eTactileKit control circuit board measuring 40mm by 60mm, housed in a red 3D-printed enclosure. From this control board, a connection extends to a human index finger shown in profile on the right side of the figure. Attached to the finger pad is a flexible orange PCB with a visible electrode array. An enlarged circular inset details the electrode configuration, showing five gold circular electrodes with measurements indicating 2.0mm center-to-center spacing and 1.4mm individual electrode diameter. Red traces connect the electrodes, demonstrating the flexible PCB routing.}
    \label{fig:ImplementationSetup}
\end{figure*}

\section{Implementation}\label{sec:implementation}
To realize TactDeform's parametric approach for emulating finger pad deformations, we implemented a modular hardware-software architecture that enables real-time spatio-temporal pattern generation. 
\rr{Building on the dual-context approach introduced in Figure~\ref{fig:concept},} this section describes the technical implementation that \rr{translates interaction and geometric contexts into electro-tactile patterns.}

\subsection{Hardware Implementation}\label{subsec:HardwareImplementation}

Our hardware configuration comprises four interconnected components \rr{(Figure~\ref{fig:ImplementationSetup})} that enable high-resolution spatio-temporal pattern delivery to the finger pad.

\emph{VR Head-Mounted Display:} A Meta Quest 3 HMD \rr{(Figure~\ref{fig:ImplementationSetup}B)} with native hand tracking (Meta Interaction SDK) captures real-time hand positions and proximity to virtual geometries. The HMD provides synchronized visual feedback while tracking finger orientation and movement, essential for our parametric approach to accurately map interaction contexts to deformation patterns.

\emph{Processing Unit:} The HMD connects to a desktop PC \rr{(Figure~\ref{fig:ImplementationSetup}A)} via Quest Link, running a Unity-based application that implements the parametric pattern generation pipeline. This configuration provides the computational resources needed for real-time parameter calculation and pattern synthesis while maintaining immersive VR.

\emph{Finger-worn Tactile Interface:} A flexible PCB \rr{(Figure~\ref{fig:ImplementationSetup}D)} with 32 electrodes attaches to the user's dominant index finger pad using breathable medical tape (3M Nexcare Gentle Paper). The electrodes \rr{are arranged in a 6$\times$6 grid layout with the four corner positions excluded, resulting in 32 active electrodes that} conform to the finger pad's natural curvature. With 1.4mm diameter electrodes and 2.0mm centre-to-centre spacing \rr{(Figure~\ref{fig:ImplementationSetup}D1)}, the array provides the spatial resolution necessary for rendering parametric deformation patterns. This spacing aligns with the two-point discrimination threshold for electro-tactile stimulation (2-4mm)~\cite{takami23extedge,Sobue2025MeasuringFingertips} and enables efficient and precise spatial pattern control. The PCB extends from the palm side with a 4.34mm width, connecting to the \rr{control circuit}, minimising interference with natural hand movement and hand tracking. The entire body-worn system weighed 46.05 grams.  

\emph{\rr{Electro-tactile} Feedback System:} We used an open-source toolkit for electro-tactile interface development~\cite{etactilekit2025}. The control hardware \rr{(Figure~\ref{fig:ImplementationSetup}C)} is housed in a 3D-printed enclosure secured to the user's forearm (sized 60 $\times$ 40 $\times$ 12mm), \rr{communicating with the desktop PC via USB serial connection at 921600 baud rate, and} connecting to the Finger-worn Tactile Interface using \rr{two 16-pin} FPC connectors, \rr{each pin connecting to the 32 electrodes. We incorporate monophasic anodic~\cite{kajimoto_electro-tactile_2016} controlled current (0-10mA) pulses with 10$\mu$A step size using its current controller. This is to enable localized stimulation while eliminating the remote (\textit{referred}) sensations occurring due to cathodic stimulation~\cite{kajimoto_electro-tactile_2016}.}
The hardware ensures constant current pulses are delivered despite the variations in skin impedance. A fast switching multiplexer (Microchip HV513) circuit enables time-division scanning to activate each electrode at a given time. We use a pulse width of 200$\mu$s with a 45$\mu$s interchannel discharge interval and a stimulation frequency of 125 Hz. These parameters align with previous work in the literature~\cite{jingu2024shaping, withana2018tacttoo}, with the exception that we reduced the interchannel discharge interval by 5$\mu$s to accommodate our 32-electrode array and ensure sufficient time for complete stimulation cycles.

\subsection{Software Architecture}
\rr{The software implementation is built in Unity 2022.3.53f1, interfacing with the control hardware through USB serial communication to deliver real-time tactile feedback synchronized with VR interactions at 72Hz.} TactDeform's software pipeline implements the parametric pattern generation and dual-context approach through real-time processing of hand tracking data.

\emph{Parametric Classification Pipeline:} The core classification component continuously analyzes hand tracking data to extract parameters for both interaction contexts (approaching, contact, sliding) and geometric contexts (faces, edges, corners, texture levels). These parameters feed into the pattern generation system that synthesizes spatio-temporal patterns based on the dual-context approach. The classification operates within Unity's frame update cycle, ensuring consistent temporal sampling for accurate parameter extraction.

\emph{Pattern Generation and Stabilization:} The pattern manager implements a finite state machine with three states: Idle (no contact), Approaching (initial contact with geometric feature parameters), and Interacting (sustained contact with orientation or texture parameters)\rr{, with transitions triggered by contact detection (Idle to Approaching), sustained contact stability (Approaching to Interacting), and contact loss (any state to Idle)}. To ensure stable parametric patterns, the system implements temporal filtering that smooths parameter fluctuations from tracking noise and hand tremor. Parameters must remain stable over brief time windows before triggering pattern updates and state transitions. This stabilization maintains perceptual continuity while preserving responsiveness to intentional parameter changes, ensuring smooth pattern evolution without abrupt changes that could disrupt the deformation emulation.

\emph{Parametric Pattern Synthesis:} The system maps extracted parameters to spatio-temporal patterns through configurable parameter-to-pattern functions\rr{, implementing the parametric formulations detailed in Section~\ref{sec:system}}. During approaching, geometric parameters determine the spatial pattern type and temporal expansion rate \rr{(Equation~\ref{eq:approach_rate})}: face parameters generate patterns that expand from the center of the electrode array\rr{;} edge parameters select from four directional line patterns (horizontal, vertical, and two diagonal orientations)\rr{, following} the orientation perception threshold \cite{bensmaia_tactile_2008}\rr{,} based on the edge's geometric orientation\rr{;} and corner parameters create concentrated point patterns at the array centre. 

During contact, the base geometric pattern is maintained while orientation parameters control the spatial shift of the stimulation region across the electrode array \rr{(Equation~\ref{eq:contact_pattern})}. This orientation mapping is limited to left-right directions, as the electrode array covers the primary contact area while the top section remains uncovered. During sliding, the base geometric pattern is combined with texture parameters through pattern multiplication \rr{(Equations~\ref{eq:texture_rate}--\ref{eq:texture_pattern})}, where texture parameters (smooth/rough/rougher) modulate the pattern shift distance (0/2/4mm) synchronized with finger movement. The system implements directional coupling where patterns rotate based on movement direction; for example, rightward finger movement triggers leftward pattern shift; emulating the relative motion between finger pad and surface.

\emph{Spatio-Temporal Synchronization:} The software maintains precise alignment between virtual geometry interactions and parametric patterns. Pattern frame advancement is tied to physical electrode spacing (2mm), ensuring that spatial patterns align with the hardware's resolution. When finger movement exceeds frame-to-frame thresholds, the system interpolates pattern parameters to maintain continuous deformation emulation even during rapid movements.

\emph{Performance Optimization:} All parameter extraction and pattern synthesis occurs within Unity's LateUpdate() cycle for consistent timing\rr{, synchronized with the 72Hz VR rendering loop}. The system uses event-driven parameter evaluation, recalculating patterns only when parameters change significantly, minimizing computational overhead while maintaining sub-frame latency \rr{($<$ 14ms)} between parameter changes and pattern updates.
\section{User Study}\label{sec:userStudy}
To evaluate TactDeform's effectiveness in rendering geometric and textural information through parametric tactile patterns, we conducted a user study examining both controlled pattern recognition and naturalistic exploration behaviours. This section first describes our pilot studies that refined the parametric patterns (Section~\ref{subsec:pilotStudy}), followed by our research questions (Section~\ref{subsec:researchQuestion}). We then present the main study design (Section~\ref{subsec:studyDesign}) with details of our two-phase study: Phase 1 validated parametric patterns through controlled tasks for geometric feature identification, contact pattern preferences, and texture discrimination (Section~\ref{subsec:Phase1Design}), while Phase 2 examined free exploration behaviours with complex 3D geometries comparing TactDeform against baseline approaches (Section~\ref{subsec:Phase2Design}).

\subsection{\rr{Iterative Design Refinement Through Pilot Studies}}\label{subsec:pilotStudy}

\rr{To validate and refine TactDeform's parametric approach, we conducted three pilot rounds with independent participant groups. Those pilots shaped our parametric pattern designs as described in Section~\ref{sec:spatio_temporal_pattern_generation}. Pilot 1 established temporal characteristics for geometric feature rendering, which refined the Equation~\ref{eq:approach_interval}, Pilot 2 examined directional encoding for texture rendering, forming the Equation~\ref{eq:texture_rate}, and Pilot 3 validated the complete system with both Pilot 1 and 2 designs.}

\rr{\textbf{Pilot 1: Pattern-Movement Synchronization (N=5).}
We first investigated how pattern timing should synchronize with finger movement during approaching interactions. Participants compared two strategies: patterns updating at fixed time intervals (0ms [static], 50ms, 100ms, 200ms) versus patterns synchronized to finger position (updating every 0.5×, 1×, or 1.5× electrode spacing, where 1× means moving 2mm advances one pattern frame). All participants preferred dynamic patterns over static, describing them as more natural. Among dynamic approaches, 80\% (4/5) preferred 1× synchronization, reporting optimal alignment between movement and sensation; smaller values felt ``too fast'' (P4) while larger ones ``lagged behind,'' (P1) and constant time intervals created speed-dependent mismatches. This established the $\alpha = 1.0$ parameter in Equations~\ref{eq:approach_rate} and \ref{eq:approach_interval}, ensuring direct spatial correspondence between finger movement and pattern progression.}

\rr{\textbf{Pilot 2: Movement Direction Resolution (N=5).}
Building on these findings, we examined how many directional levels are needed to encode sliding movement. Each participant completed three comparisons, testing whether 4-direction, 8-direction, or continuous direction tracking improved texture perception compared to static non-directional patterns. We found that 8-direction quantization (forward, backward, left, right, plus four diagonal corners) balanced perceptual quality: 4 directions caused ``jumpy'' (P1) transitions during diagonal movement, continuous tracking created ``buzzing'' (P2) from micro-variations, static patterns feel ``constant touching'' (P5) and ``not moving'' (P1) while 8 directions provided smooth yet stable feedback. This established the directional encoding for texture patterns (Equation~\ref{eq:texture_rate} and \ref{eq:texture_pattern}), where the shift direction adapts to finger movement direction across the surface.}

\rr{\textbf{Pilot 3: Complete System Validation (N=7).}
Before the formal study, we tested the complete system through our three main tasks (geometric feature identification, contact pattern preference, texture discrimination) alongside comparison tasks that validated Pilots 1-2 design choices. Participants' high accuracy and qualitative feedback confirmed that velocity-synchronized patterns (from Pilot 1) combined with 8-direction texture encoding (from Pilot 2) created coherent tactile experiences matching natural finger pad deformations. This validation enabled us to streamline the formal study by removing temporal parameter comparison tasks from Phase 1 and finalizing the Phase 2 protocol, reducing session time while maintaining focus on questions requiring larger sample sizes.}

\subsection{Research Questions}\label{subsec:researchQuestion}
We formulated three research questions to guide the formal evaluation:

\begin{itemize}
\item \textbf{RQ1}: How effectively can TactDeform's parametric patterns convey geometric and textural information across different interaction contexts?
\item \textbf{RQ2}: \rr{How does TactDeform's parametric approach shape user exploration behaviors and preferences compared to alternative rendering approaches?}
\item \textbf{RQ3}: \rr{How do objects ranging from simple geometric primitives to complex organic shapes reveal TactDeform's capabilities and limitations?}
\end{itemize}

\subsection{Study Design}\label{subsec:studyDesign}

\begin{figure*}
    \centering
    \includegraphics[width=1\linewidth]{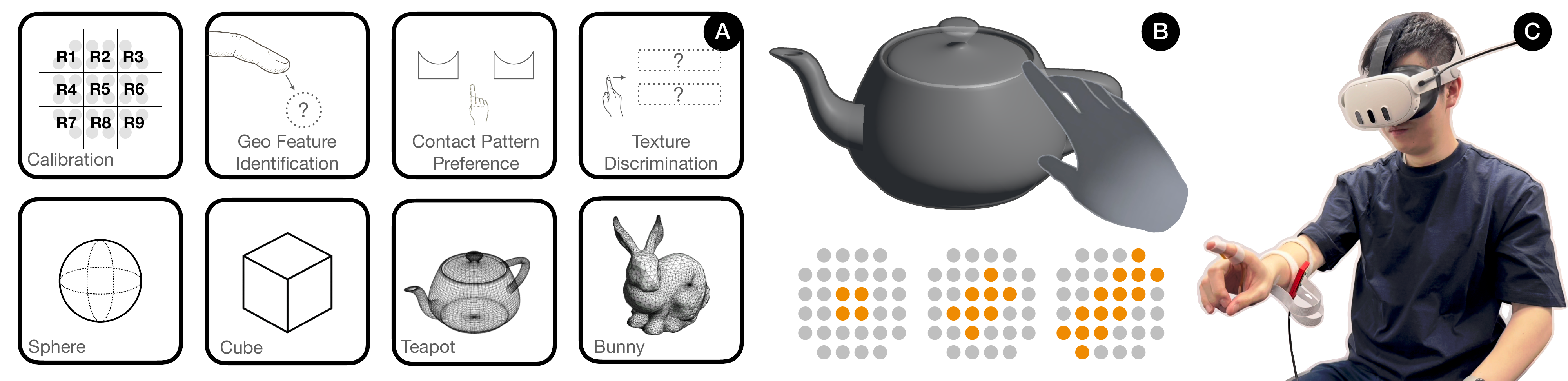}
    \caption{User Study Design Overview. (A) Overview of the two-phase study' procedure, validating TactDeform's concept through controlled tasks (Phase 1) and free exploration (Phase 2). (B) Electrode activation patterns showing how TactDeform renders different geometric features through spatial stimulation on the 32-electrode array. (C) A participant experiencing TactDeform during the study, wearing the Meta Quest 3 HMD with the finger-worn tactile interface attached to their index finger.}
    \Description{Figure showing the user study design in three parts. Part A displays the study structure as two rows of boxes: the top row shows Phase 1's three tasks (Calibration with a 3x3 grid labelled R1-R9, Geometric Feature Identification with dotted patterns representing faces/edges/corners, Contact Pattern Preference with angle diagrams, and Texture Discrimination with different line patterns), while the bottom row shows Phase 2's four test objects (wireframe sphere, cube, teapot, and bunny models) of increasing geometric complexity. Part B shows a rendered 3D teapot model above three 6x6 electrode arrays demonstrating different activation patterns - the arrays show orange-highlighted electrodes forming different spatial patterns for rendering geometric features. Part C is a photograph of a study participant wearing a Meta Quest 3 VR headset and the TactDeform system, with the finger-worn tactile interface visible on their index finger connected to a control unit on their forearm.}
    \label{fig:userStudy}
\end{figure*}

To address these research questions, we conducted a formal user study \rr{(N=24)} comprising two phases \rr{(Figure~\ref{fig:userStudy}A)}. Phase 1 employed controlled tasks to validate the discriminability and effectiveness of our parametric patterns (addressing RQ1). Phase 2 examined naturalistic exploration behaviours with complex 3D geometries, comparing TactDeform against baseline rendering approaches (addressing RQ2 and RQ3).

\textbf{Participants}: We recruited 26 participants ( \rr{12 women, 14 men}, age 18-44) through word of mouth. \rr{Participants underwent eligibility screening with the following exclusion criteria: (1) cardiac pacemaker or other implanted electronic medical devices, (2) pregnancy, (3) history of epilepsy or seizures, (4) skin conditions or injuries on the dominant index finger that could interfere with electrode placement, (5) peripheral neuropathy or reduced tactile sensitivity in the hands, and (6) severe motion sickness or vestibular disorders that could be triggered by VR. These criteria align with established safety protocols for electro-tactile stimulation~\cite{kajimoto_electro-tactile_2016} and VR studies~\cite{siu_virtual_2020}.} Participant's prior experience varied: \rr{18} had used VR before (\rr{13} rarely, \rr{3} monthly, \rr{2} weekly, primarily for gaming), \rr{19} had haptic device experience, and \rr{6} had used electro-tactile devices, with only \rr{1} having combined electro-tactile-VR experience. Two participants were excluded from analysis: P06 due to performance significantly below chance level in geometric feature identification (0.238 vs. 0.333 chance) and texture discrimination (0.5 vs. 0.5 chance), and P08 due to incomplete study completion. This resulted in \rr{24} participants for analysis (\rr{12} women, \rr{12} men). The study was approved by the university ethics committee (2024/HE001103).

\textbf{Apparatus}: The study used the TactDeform implementation described in Section \ref{sec:implementation}, with the finger-worn interface attached to participants' dominant index finger \rr{(Figure~\ref{fig:userStudy}C)}. The study was conducted in a controlled lab environment.

\textbf{Procedure}: \rr{The complete session lasted approximately 2.5 hours, including 15 minutes of introduction, 30 minutes of calibration, and the remaining time for experimental tasks.} Participants remained seated throughout. Instructions were standardized, with no specific guidance about exploration strategies. \rr{Unlike prior electro-tactile studies that require training sessions~\cite{suga_3d_2023, suga_presentation_2024}, we provided no practice trials or trainning sessions to assess whether our patterns are intuitively interpretable.} Participants could request breaks between tasks. No compensation was provided.

\textbf{Calibration}: Calibration of electro-tactile feedback interfaces needs careful attention when used in spatial rendering applications due to individual differences in skin properties and nerve sensitivity at different locations lead to variability in perceptual thresholds~\cite{Sobue_Optimal_2025,duente_shock_2024}.
Following established protocols~\cite{kajimoto_electro-tactile_2016}, we calibrated pulse amplitudes before each user study session. \rr{Pulse amplitude was the only parameter calibrated for individual participants. All other electro-tactile parameters, including pulse width, frequency, and inter-channel discharge interval, remained constant across all studies as specified in Section~\ref{subsec:HardwareImplementation}. To reduce the time consumed to calibrate all 32 electrode locations, the 32 electrodes were grouped into 9 regions (Figure~\ref{fig:userStudy}A1) for regional calibration.}
Starting from the centre region, pulse amplitude was gradually increased until participants experienced discomfort, then decreased until they perceived clear, pain-free stimulation. The remaining regions were matched to the centre's perceived intensity to ensure a consistent sensation across the finger pad. Finally, all regions were activated simultaneously for global adjustment. No study patterns were revealed during calibration. 

\subsection{Phase 1: Pattern Validation}\label{subsec:Phase1Design}
Phase 1 comprised three tasks \rr{(Figure~\ref{fig:userStudy}A)} validating the spatio-temporal patterns used in Phase 2's free exploration. Tasks followed a fixed sequence matching natural VR interaction progression: approaching objects, maintaining contact, then sliding across surfaces. The tasks design and procedure are inspired by existing haptic perception study \cite{jones_application_2013, bensmaia_tactile_2008}.

Phase 1 specifically addressed research question RQ1. We further break down this research question into three areas: geometric feature discrimination during approaching interactions (RQ1a), pattern preferences during sustained contact (RQ1b), and texture discrimination during sliding interactions (RQ1c).

\textbf{Post-Task Interviews}: After each task, we conducted semi-structured interviews to gather qualitative feedback. Following established practices \cite{jingu2024shaping, strohmeier_barefoot_2020, kildal_3d-press_2010, friedman_magnitude_2008}, we evaluated participants' subjective experiences, which is an effective approach for investigating haptic experiences that are difficult to assess through other methods \cite{strohmeier_pulse_2018}. All interviews were audio-recorded and transcribed.

\subsubsection{Task 1: Geometric Feature Identification}\label{subsubsec:Phase1DesignT1}

\textbf{Goal}: Validate users' ability to distinguish geometric features through parametric deformation patterns during approaching interactions (RQ1a).
\textbf{Setup}: Participants faced an invisible virtual object positioned 45cm away and 20cm below eye level, touchable within a 5cm radius sphere. A visual indicator marked the location, disappearing when the finger approached within 8cm to eliminate visual bias.
\textbf{Procedure}: Using 3-alternative forced choice (3AFC), participants identified ``Which geometric feature do you perceive?" with options: face, edge, or corner. To experience the approaching context, participants started beyond 8cm and approached the invisible object. The geometric feature was always tangent to the index finger, eliminating position confounds. Participants could approach multiple times without time limits but had to withdraw beyond 8cm between approaches.
\textbf{Parametric Patterns}: The system delivered deformation-emulating patterns as described in Section 3 \rr{and shown in Figure~\ref{fig:teaser}.1}: face patterns used diverging rings expanding from the finger pad centre, edge patterns employed directional linear stimulation, and corner patterns created concentrated point stimulation.
\textbf{Conditions}: 21 trials with 7 repetitions each of face, edge, and corner patterns, randomized. Edge patterns included 4 orientations (horizontal, vertical, diagonal slopes); to balance the design, three edge orientations were presented twice and one orientation was presented once per session, with the single-presentation orientation randomized across participants. Face and corner patterns used single parametric representations.
\textbf{Measures}: 3AFC responses, confidence ratings (5-point Likert), approaching counts, and trial duration.

\subsubsection{Task 2: Contact Pattern Preference}\label{subsubsec:Phase1DesignT2}
\textbf{Goal}: Understand preferences for orientation-specific versus uniform patterns during contact interactions (RQ1b).
\textbf{Setup}: Two visible objects (20$\times$5$\times$5cm) were placed 40cm away and 20cm below eye level, with one 20cm left and the other 20cm right of the centre eye position. One delivered orientation-specific patterns adapting to contact angle, the other provided uniform stimulation.
\textbf{Procedure}: Participants freely explored both objects through stationary contact. Stimulation occurred only when the finger pad remained stationary; movement immediately stopped stimulation. After exploration, participants answered ``Which tactile feedback feels better?'' (2AFC) with confidence rating.
\textbf{Parametric Patterns}: The orientation-specific object implemented four parametric levels (steep left, shallow left, shallow right, steep right)  \rr{as shown in Figure~\ref{fig:teaser}.2}, shifting stimulation across the finger pad to emulate natural contact deformation at different angles.
\textbf{Conditions}: 14 trials with counterbalanced pattern-to-position assignment and randomized trial order.
\textbf{Measures}: 2AFC responses, confidence ratings, interaction angles, interaction counts/durations, and trial duration.

\subsubsection{Task 3: Texture Discrimination}\label{subsubsec:Phase1DesignT3}
\textbf{Goal}: Assess texture discrimination through parametric shift mechanisms during sliding interactions (RQ1c).
\textbf{Setup}: Two invisible surfaces placed 40cm away, one 10cm below and another 25cm below eye level, with 30cm horizontal extent.
\textbf{Procedure}: Participants selected a surface and moved rightward from start to end. The system delivered patterns only during rightward movement to isolate texture perception. After exploring both surfaces, participants answered ``Which surface feels rougher?" (2AFC) with confidence rating.
\textbf{Parametric Patterns}: Three texture levels through shift parameters \rr{as shown in Figure~\ref{fig:teaser}.3}: Level 1 (smooth) with zero shift, Level 2 (rough) with one electrode spacing shift (2mm), Level 3 (rougher) with two electrode spacing shift (4mm).
\textbf{Conditions}: 24 trials presenting all level combinations, balanced for pairing and position.
\textbf{Measures}: 2AFC responses, confidence ratings, interaction counts, and trial duration.

\subsection{Phase 2: Free 3D Geometry Exploration} \label{subsec:Phase2Design}
\textbf{Goal}: Phase 2 investigates how users \rr{respond to TactDeform's parametric approach compared to alternative rendering approaches, and objects of varying complexity,} during unconstrained exploration, \rr{examining exploration behaviors and preferences (RQ2), and how objects ranging from simple to complex shapes reveal TactDeform's capabilities (RQ3).}

\noindent \textbf{Tactile Rendering Approaches}: The phase compared three tactile rendering approaches representing different levels of sophistication in spatial and temporal feedback:
\begin{itemize}
\item \textbf{Uniform Activation (UA)}: Uniform electrode activation when finger enters object boundary. UA represents the simplest binary feedback approach common in vibrotactile systems, where contact triggers uniform stimulation regardless of geometric features or interaction context.
\item \textbf{Contact-Area Mapping (CAM)}: Individual electrode activation based on localized contact detection. This baseline represents direct spatial mapping approaches used in current electro-tactile interfaces~\rr{\cite{Teng_Seeing_2025, jingu2024shaping}}, where stimulation directly corresponds to contact area without temporal dynamics or contextual adaptation.
\item \textbf{TactDeform}: Full parametric system with dual-context approach (interaction and geometric contexts) emulating finger pad deformations. This approach adds spatio-temporal patterns and context-aware adaptation to evaluate the contribution of parametric deformation emulation.
\end{itemize}

\noindent \textbf{Objects}: Four objects with increasing geometric complexity: sphere, cube, teapot, and bunny (20$\times$20$\times$20cm bounding dimensions) as shown in Figure~\ref{fig:userStudy}A, representing progression from simple primitives to complex organic forms.

\noindent \textbf{Procedure}: Participants freely explored each object with each rendering approach (12 trials total), with counterbalanced ordering. Participants were encouraged to think aloud during exploration and had unlimited time and interaction freedom to explore each virtual object.

\noindent \textbf{Post-Phase Interview}: Semi-structured interviews were conducted to gain knowledge about the experiences with the three different tactile feedback approaches across object types and exploration strategies. \rr{Phase 2 exploration and post-phase interviews were audio recorded.}
\section{Results}\label{sec:results}

This section presents the \rr{results from our two-phase empirical evaluation of TactDeform with 24 participants.} Phase 1 validated the fundamental parametric patterns through three controlled tasks (RQ1a-c), while Phase 2 examined naturalistic exploration with complex 3D geometries (RQ2, RQ3).

All statistical analyses were conducted using Python (version 3.13.2) with statsmodels and pingouin packages. We employed repeated measures ANOVA for within-subjects comparisons, accounting for the fact that each participant experienced all conditions. \rr{Post hoc} pairwise comparisons used Tukey HSD correction for multiple comparisons. Statistical significance was set at $\alpha = 0.05$, with Bonferroni correction applied where appropriate for multiple factor comparisons. Qualitative data were analyzed using reflexive thematic analysis~\cite{braun_using_2006}, with two researchers independently coding transcribed interviews to ensure reliability.

\subsection{Phase 1: Validating Parametric Patterns}

Phase 1 systematically evaluated the core parametric design principles underlying TactDeform's tactile rendering approach across three \rr{interaction contexts: Approaching (Task 1, Section~\ref{subsec:task1}), Contact (Task 2, Section~\ref{subsec:task2}), and Sliding (Task 3, Section~\ref{subsec:task3}). Addressing RQ1: How effectively can TactDeform’s parametric patterns convey geometric and textural information across different interaction contexts?} 
Each task presents quantitative results, qualitative insights from user interviews, and targeted discussions of implications for parametric tactile design.

\begin{figure*}
    \centering
    \includegraphics[width=1\linewidth]{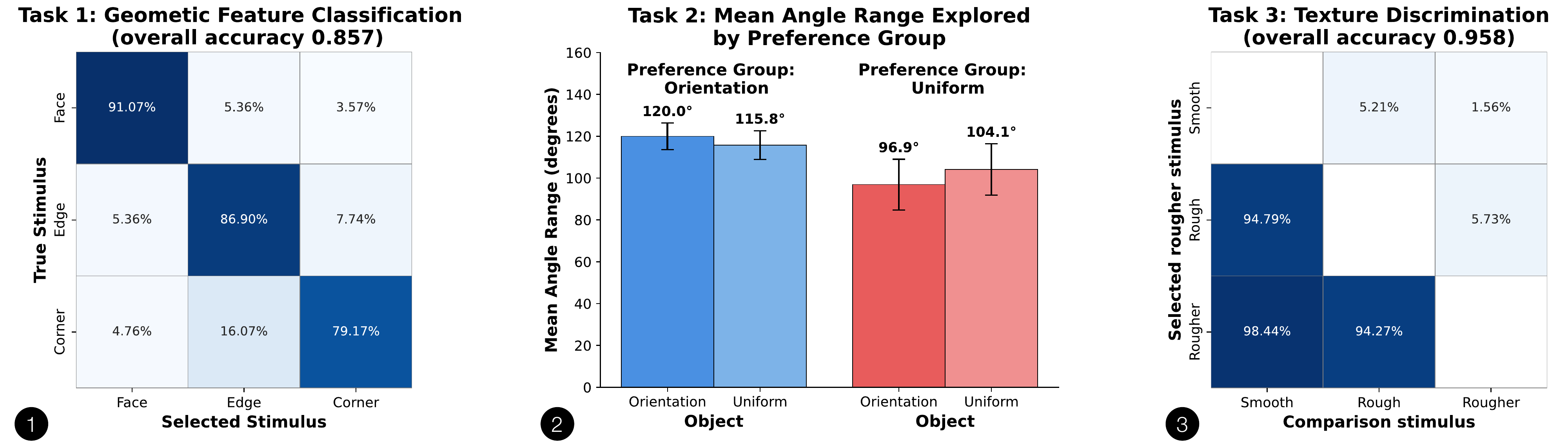}
    \caption{Phase 1 Results.
    (1) Confusion matrix for Task 1 geometric feature classification. Participants classified haptic stimuli into Face, Edge, or Corner categories. Values show the percentage of trials where the true stimulus (row) was classified as the selected category (column). Overall accuracy was \rr{0.857} across all participants.
    (2) Mean angle range explored by preference groups in Task 2. Participants were categorized into Orientation-preference and Uniform-preference groups based on their preferred patterns. Bars show mean angle range (in degrees) with standard error for each group-object combination. Blue bars represent the Orientation group, pink bars represent the Uniform group.
    (3) Pairwise preference matrix for Task 3 texture discrimination. The matrix shows the probability that the row stimulus was chosen as rougher than the column stimulus. Stimuli are ordered by increasing roughness: Lv1 \textit{Smooth}, Lv2 \textit{Rough}, Lv3 \textit{Rougher}. Values represent percentages with overall accuracy of \rr{0.958}.}
    \Description{A three-panel figure displaying Phase 1 experimental results. 
    Left panel: A confusion matrix presents the results of the geometric feature classification task, where participants identified whether tactile stimuli represented faces, edges, or corners of virtual objects. The matrix shows strong diagonal performance, indicating good classification accuracy: faces were correctly identified 91.07\% of the time, edges 86.9\%, and corners 79.17\%. The most common confusion occurred between corners and edges (16.07\% of corners misclassified as edges), suggesting that distinguishing sharp geometric transitions presents the greatest challenge. The overall accuracy of 85.7\% demonstrates that users can effectively discriminate between fundamental geometric features through tactile exploration, though corner detection remains more challenging than face or edge identification.
    Middle panel: A bar chart analyzes exploration behavior by comparing the angular range of finger movement between two preference groups identified in our study. Participants were categorized based on their interaction patterns: those with orientation preferences and those with uniform preferences. The results reveal distinct behavioral differences between groups. The orientation preference group demonstrated significantly broader exploration ranges, covering 120.0 degrees and 115.8 degrees for different object types, while the uniform preference group showed more constrained exploration patterns with ranges of 96.9 degrees and 104.1 degrees. This finding suggests that individual differences in tactile exploration strategies significantly impact how users interact with virtual objects, with some users employing more comprehensive scanning behaviours while others use more focused, localized exploration patterns.
    Right panel: A confusion matrix evaluates participants' performance in texture discrimination, where they identified which of two stimuli felt rougher. The task achieved exceptionally high accuracy (95.8\%), demonstrating users' strong ability to discriminate tactile texture through the virtual system. The matrix shows excellent performance across all texture comparisons: when comparing smooth vs. rough textures, participants correctly identified the rough stimulus 94.79\% of the time; when comparing rough vs. rougher textures, accuracy was 94.27\%. The minimal off-diagonal values (1.56-5.73\%) indicate very few misclassifications, suggesting that the tactile rendering system effectively conveys texture information and that participants can reliably distinguish between different roughness levels through virtual touch.}
    \label{fig:phase1Results}
\end{figure*}

\subsection{Task 1: Geometric Feature Recognition (RQ1a)}\label{subsec:task1}

\rr{We evaluated TactDeform's ability to convey geometric features (faces, edges, corners) through parametric patterns during approaching interactions, addressing RQ1a. For detailed protocol and setup, see Section~\ref{subsubsec:Phase1DesignT1}.}

\subsubsection{Quantitative Results:}
\rr{Participants achieved high overall accuracy in geometric feature recognition ($M = 85.71\%$, $SD = 10.41\%$, $N = 24$). A repeated measures ANOVA found significant differences between feature types ($F(2, 46) = 3.64$, $p = .031$, $\eta^2_p = .16$). Descriptively, faces showed the highest mean accuracy ($M = 91.07\%$, $SD = 12.51\%$), followed by edges ($M = 86.90\%$, $SD = 13.92\%$), and corners ($M = 79.17\%$, $SD = 19.30\%$) as shown in Figure~\ref{fig:phase1Results}.1. The lower accuracy for corners suggests potential for improving the parametric representation of concentrated point features.}

\rr{Exploration behavior algins with the accuracy, varied by feature, with corners requiring more replays, i.e., approaching interaction ($M = 5.12$, $SD = 3.21$) than edges ($M = 4.48$, $SD = 2.87$, $p = .012$) or faces ($M = 3.57$, $SD = 2.41$, $p < .001$). Accuracy showed a weak ($r = .192$) but significant learning effect across trials ($p < .001$), increasing from 75.0\% in early trials to 89.9\% in later trials. Confidence ratings ($M = 4.04$, $SD = 0.65$ on a 5-point scale) correlated moderately with accuracy ($r = .267$, $p < .001$)}

\subsubsection{Qualitative Insights:}
\rr{Participants consistently adopted spatial categorization strategies without instruction, describing edges as \textit{``lines,''} corners as \textit{``points,''} and faces as \textit{``broad areas''} across interviews. Participants explicitly noted the learning process: \textit{``first 3 tasks... pretty much guessing... then I knew the pattern''} (P03) and \textit{``after 5 to 6 times I learned how to distinguish''} (P12), indicating that parametric patterns align with natural tactile processing strategies.}

\subsection{Task 2: Contact Pattern Preference (RQ1b)}\label{subsec:task2}

\rr{This task evaluated which contact pattern approach (orientation-specific versus uniform) better supports stationary interaction, as measured by user preferences, addressing RQ1b. For detailed protocol and setup, see Section~\ref{subsubsec:Phase1DesignT2}.}

\subsubsection{Quantitative Results}
Participants exhibited different preference distribution: \rr{14 out of 24 participants (58.3\%) strongly preferred orientation-specific patterns (selecting $\geq 70\%$ of the orientation-specific patterns), while 41.7\% (10/24) strongly preferred uniform patterns (selecting $\geq 70\%$ of the uniform patterns). Remarkably, no participants exhibited mixed preferences (31-69\% range), suggesting fundamentally distinct tactile processing strategies rather than a preference continuum.}

\rr{Analysis of preference groups revealed that participants who prefer orientation-specific patterns exhibited wider angular exploration regardless of object type (orientation: $M = 120.0\degree$, uniform: $M = 115.8\degree$) compared to participants who prefers uniform patterns (orientation: $M = 96.9\degree$, uniform: $M = 104.1\degree$) as shown in Figure~\ref{fig:phase1Results}.2. The difference was most pronounced for orientation objects ($23.1\degree$ difference, $p = .081$, $d = 0.76$). Within-group comparisons showed no significant preference-based modulation of exploration (both $p > .27$), suggesting that stated preferences reflect perceptual quality judgments rather than behavioural engagement patterns.}

\subsubsection{Qualitative Insights}
\rr{Interview analysis validated this further; it revealed two contrasting judgment strategies. 11 out of 24 participants preferred the object that provides a stronger sensation; they consistently select \textit{``the stronger one''} for clarity (P03) or seek \textit{``stronger intensity in general''} (P09). In contrast, 11 out of 24 participants valued spatial specificity, linking localized activation to realism: \textit{``only the left side is vibrating... more natural''} (P02) and \textit{``just turn on the contacted surface is more realistic''} (P10). Three participants discovered the orientation-specific system's capabilities during exploration, shifting from passive reception to active investigation, exemplified by P02's evolution from preferring \textit{``strong stimulation''} to \textit{``changing stimulation [that] would actually suit the surface more.''}} \rr{Both qualitative and quantitative results indicate two distinct user populations with different haptic judgment methods, suggesting that adaptive or customizable tactile rendering systems may better accommodate individual differences, aligning with the recent research~\cite{kim_defining_2020}.}

\subsection{Task 3: Texture Discrimination (RQ1c)}\label{subsec:task3}

\rr{Task 3 evaluated TactDeform's ability to render distinguishable texture levels (smooth, rough, rougher) through parametric shifting patterns during sliding interactions, addressing RQ1c. For detailed protocol and setup, see Section~\ref{subsubsec:Phase1DesignT3}.}

\subsubsection{Quantitative Results}

\rr{Texture discrimination achieved high accuracy overall ($M = 95.83\%$, $SD = 5.07\%$, $N = 24$) (Figure \ref{fig:phase1Results}.3). Mean accuracy ranged from 94.27\% (rough vs rougher) to 98.44\% (smooth vs rougher), though these descriptive differences were not statistically significant ($F(2, 46) = 1.80$, $p = .174$), suggesting near-ceiling 
performance across all texture pairs.}

Participants demonstrated \rr{strong awareness of the texture.} Confidence ratings strongly predicted actual performance, both at the trial level (correct trials: $M = \rr{4.14}$ vs incorrect: $M = \rr{3.39}$; $t(\rr{574}) = \rr{6.22}$, $p < .001$, $d = 1.08$) and individual level ($\rr{r = .748}$, $p < .001$). This awareness manifested behaviorally through adaptive sampling: participants averaged \rr{2.41} replays per trial, but adjusted their exploration based on difficulty. The challenging rough vs rougher comparison required significantly more replays (\rr{2.82}) than the easier smooth vs rougher comparison (\rr{1.87}; $U = \rr{11096.5}$, $p < .001$), with replay counts inversely correlating with confidence ($r = \rr{-.469}$, $p < .001$), \rr{indicating genuine perceptual understanding rather than guessing strategies.}

\subsubsection{Qualitative Insights}
\rr{Participants spontaneously developed multi-dimensional spatial-temporal exploration strategies for texture perception, naturally decoding TactDeform's parametric shift mechanism through coordinated perceptual channels. Most participants simultaneously integrated temporal patterns (54\% describing \textit{``alternations''} or \textit{``wave-like patterns''}) and spatial coverage variations (50\% noting \textit{``different points of my finger''} activation). Participants described the perceived texture as material metaphors: \textit{``concrete with tiny rocks''} (P03) for roughest surfaces, \textit{``sparkling bubbles popping''} (P12) for medium roughness, and \textit{``table surface''} (P13) for smooth textures. Multiple participants independently found that movement speed modulated perception clarity (\textit{``I can feel it more clearly if I go slower,''} P14), revealing how users naturally optimize exploration strategies through our parametric shift mechanism.}
\subsection{Phase 2: Free Exploration with Complex Geometries (RQ2, RQ3)}

\rr{Phase 2 evaluated how TactDeform's parametric approach supports natural exploration compared to baseline rendering approaches (Uniform Activation, Contact-Area Mapping) across objects of increasing geometric complexity (sphere, cube, teapot, bunny). This phase addressed RQ2 (how TactDeform's parametric approach shapes user exploration behaviours and preferences compared to alternative rendering approaches) and RQ3 (how objects ranging from simple geometric primitives to complex organic shapes reveal TactDeform's capabilities and limitations). For rendering approach description, detailed protocol and setup, see Section~\ref{subsec:Phase2Design}.}

\rr{We analysed 288 interactions from 24 participants exploring four objects with three rendering approaches, using think-aloud protocols and semi-structured interviews to identify preferences and exploration strategies.}

\subsubsection{Rendering Approach Comparison (RQ2)}

Participants' preferences revealed \rr{clear patterns across rendering approaches.} \rr{No participants preferred UA, with participants describing it as \textit{``general feedback''}} (P21) \rr{that was} \textit{``not matching to the visual cue''} (P21), and it was explicitly rejected by \rr{58\%} of participants (\rr{14/24}), who characterized its uniform activation as \textit{``always-on,''} \textit{``same everywhere,''} with \textit{``no distinction.''} This indicates that the lack of spatial characteristics affected the interactions.

Between spatially-selective approaches, \rr{two-thirds} (\rr{16/24}) preferred TactDeform's surface rendering while \rr{the remainder} (\rr{8/24}) favoured CAM's continuous volume feedback. \rr{This split revealed two distinct interaction strategies. Participants prefer CAM explore the feedback at boundaries, they poke in and out repeatedly and slowly, doing approaching interaction, as P03 noted:} \textit{``I only feel it when I touch the surface... makes me feel more like I'm touching an object.''} \rr{In contrast, navigation-oriented users valued continuous spatial awareness, they trace the object by sliding on the object surface, textures are then rendered by TactDeform.}

\rr{These preferences manifested in different exploration behaviours. More than} half of participants (\rr{58\%}, \rr{14/24}) explicitly evaluated systems based on edge and corner clarity, \rr{deliberately} seeking \textit{``line-like''} and \textit{``point-like''} sensations to verify rendering quality. TactDeform consistently received praise for crisp feature transitions, with P11 noting it \textit{``exactly reflect[s] the objects... face or the corners,''} while CAM \rr{required slower, more deliberate movements to perceive} area-based changes. \rr{UA provided no geometric information:} \textit{``The face edge and corner all feels the same''} (P25).

\subsubsection{Object Complexity Effects (RQ3)}

\rr{The four objects served as progressive diagnostic tools. The sphere's uniform curvature provided a baseline where all approaches converged to basic contact detection:} \textit{``For the sphere... I don't feel a big difference between the 3 conditions''} (P01).

The cube emerged as \rr{the} canonical test for geometric feature sensitivity, with \rr{58\%} of participants (\rr{14/24}) systematically probing for edges and corners. \rr{Performance separated clearly: TactDeform delivered distinct face/edge/corner transitions, CAM provided area-based differentiation, while UA failed completely.}

\rr{The teapot revealed differential performance on fine features. With UA, participants could not perceive detailed structures:} \textit{``I couldn't feel the tip of the spout''} (P05)\rr{, as uniform activation provided no spatial differentiation. TactDeform and CAM conveyed major edges like the rim and handle, though the entire spout structure remained challenging across all approaches.}

The bunny generated the most differentiated responses \rr{through its textured surface. TactDeform's parametric texture rendering, the only textured object, was recognized by 54\% of participants (13/24):} \textit{``a furry bunny''} (P25)\rr{,} \textit{``in some moment I can really feel I am touching a bunny''} (P26). \rr{UA remained uniformly uninformative.}
\section{Discussion}

Our evaluation of TactDeform reveals several broader takeaways that extend beyond the immediate findings to inform the future of haptic interface design and electro-tactile feedback systems.

This work demonstrates that parametric approaches to tactile rendering can successfully bridge the gap \rr{of physically restricted kinaesthetic feedback devices} and perceptual fidelity. By encoding geometric \rr{feature and texture} properties through electro-tactile patterns, TactDeform validates a principled approach to haptic design that moves beyond ad hoc pattern generation \cite{torres_hapticprint_2015, bau_teslatouch_2010}. The adoption of spatial context-based classification strategies across participants suggests that our parametric encoding aligns with fundamental perceptual mechanisms, providing a theoretical foundation for future haptic rendering systems \cite{mania_cognitive_2010, okamoto_psychophysical_2013}.

\rr{The geometric feature recognition results (Task 1, 85.71\% accuracy) validate that our parametric patterns can effectively convey fundamental spatial properties during approaching interactions. The performance hierarchy, faces (91.07\%) > edges (86.90\%) > corners (79.17\%), reveals an important design consideration: while broader spatial patterns are immediately intuitive, concentrated point features require more refined parametric representations. The spontaneous emergence of spatial categorization strategies across all participants, who consistently described features as ``broad areas,'' ``lines,'' and ``points'' without instruction, demonstrates that our deformation-to-stimulation mapping aligns with natural tactile processing mechanisms~\cite{jones_application_2013}. The significant learning effect (75.0\% to 89.9\% across trials) combined with moderate confidence-accuracy correlation ($r = .267$) suggests that parametric patterns support rapid perceptual learning, enabling users to develop genuine spatial understanding rather than memorizing arbitrary mappings. This finding is particularly important for practical applications, as it indicates that TactDeform's approach requires minimal training while maintaining interpretability, which is important for real-world deployment in VR environments.}

\rr{The exceptional texture discrimination accuracy (95.83\%) validates TactDeform's parametric shift mechanism as an effective approach for conveying surface properties through electro-tactile feedback. The success of this approach stems from leveraging multiple perceptual dimensions, including intensity, temporal patterns, and spatial coverage, rather than a single cue. This is consistent with findings that natural texture perception integrates both spatial and temporal information from mechanoreceptor populations \cite{weber_spatial_2013}.}

Furthermore, the consistent emergence of individual differences across tasks, such as different contact pattern preferences and diverse face perception profiles, \rr{aligns with the idea that effective haptic interfaces should provide customized experiences~\cite{kim_defining_2020}.} Rather than viewing these differences as obstacles, our findings suggest they represent design opportunities that systems should accommodate through adaptive rendering approaches~\cite{malvezzi_design_2021}. The strong coupling between preferences and exploration behaviors indicates that these differences reflect genuine perceptual strategies, supporting recent calls for personalized haptic design~\cite{seifi_exploiting_2017}. This insight has broader implications for the field, suggesting that future haptic systems should prioritize adaptability and user modeling over one-size-fits-all approaches.

These findings have implications for extended reality and accessibility applications. The high accuracy achieved without extensive training (\rr{85.7}\% geometric recognition, \rr{95.8}\% texture discrimination) \rr{suggests potential} for accessible interfaces supporting visually impaired users in XR environments \cite{Teng_Seeing_2025, siu_virtual_2020}. The intuitive nature of parametric patterns, validated through universal spatial extent mapping strategies, indicates that these approaches could provide effective shape communication for users who rely primarily on haptic feedback for spatial understanding.

The feature-sensitive nature of our approach, combined with the progressive complexity effects observed across different object types, also suggests promising applications in medical and industrial training where spatial feature recognition is critical \cite{motaharifar_applications_2021, leleve_haptic_2020}. The ability to emphasize edges, corners, and surface transitions could enhance training for surgical procedures, manufacturing quality control, or architectural design evaluation, where subtle geometric features often carry critical information.
\section{Limitations and Future Directions}

Our implementation presents several technical constraints that highlight directions for future development. \rr{First, we explored only monophasic anodic current; investigating cathodic stimulation could reveal different perceptual qualities or improved comfort profiles. Second,} the electrode array's limited fingertip coverage restricts orientation detection to left-right variations, preventing full 3D orientation-dependent rendering \cite{Sobue_Optimal_2025}. \rr{Finally,} our evaluation focused on single-finger interactions, while real-world haptic exploration typically involves multi-finger coordination and bi-manual strategies that could significantly enhance spatial understanding and feature detection \cite{tanaka_full-hand_2023, schorr_fingertip_2017, jingu_double-sided_2023}.

The geometric scope of our evaluation, while systematically varied in complexity from simple primitives to organic forms, represents only a limited subset of real-world shapes and interaction scenarios. Future research should evaluate performance across broader geometric categories, including deformable objects, moving targets, and environments with multiple simultaneous objects. The fine details limitations revealed by complex objects like the teapot and bunny highlight the need for more sophisticated inside-outside classification algorithms handling negative space effectively.

Perhaps most promising for future development is the potential for dynamic rendering systems that intelligently adapt based on real-time exploration context. Our interaction distance and movement findings suggest that systems could monitor finger position, movement velocity, and contact duration to optimize rendering approaches dynamically, providing the right type of feedback at the right moment while accommodating individual exploration preferences and task-specific requirements. Such systems could dynamically transition between contact-based rendering when users employ finger pad contact for surface exploration and context-based rendering when using fingertip edges for boundary detection, while adapting to in-and-out movements during boundary exploration to provide optimal spatial feedback that maximizes both geometric understanding and individual user preferences.

Finally, future work should investigate how parametric tactile patterns integrate with other haptic modalities and sensory channels in realistic application contexts. Cross-modal integration with force feedback, vibrotactile systems, and visual-auditory information remains largely unexplored, yet understanding these interactions is crucial for deploying haptic systems in complex environments where tactile feedback competes with other sensory modalities for user attention \cite{shrestha_virtual_2025, yun_real-time_2025}.
\section{Conclusion}
We presented TactDeform, a parametric approach for rendering spatio-temporal \rr{electro-tactile} patterns that emulate natural finger pad deformations during 3D geometry exploration in virtual reality. \rr{By dynamically adapting patterns to both interaction contexts (approaching, contact, sliding) and geometric contexts (features and textures), TactDeform demonstrates that electro-tactile feedback alone can convey} fine-grained spatial characteristics\rr{, without requiring force feedback apparatus, enabling large exploration range without physical constraints}. Our user study \rr{(N=24)} \rr{validates this approach through high} geometric feature identification \rr{(85.7\%} accuracy) and texture discrimination \rr{(95.8\%} accuracy), \rr{establishing} that parametric encoding of deformation signatures \rr{effectively conveys} rich geometric and textural information \rr{through electro-tactile stimulation}. \rr{Beyond these quantitative results, our findings reveal important qualitative insights: 58\% of participants spontaneously developed edge-probing exploration strategies, and individual differences in tactile processing emerged naturally, indicating that our deformation-based encoding aligns with natural perception mechanisms while highlighting the need for adaptive haptic system design.} The parametric principles underlying TactDeform extend beyond electro-tactile interfaces\rr{, as} the deformation-emulation concept applies to other haptic modalities including high-density \rr{vibrotactile} arrays and multi-finger interfaces, where understanding characteristic deformation patterns can inform rendering strategies. With broad implications spanning medical training, accessibility technologies, and interactive VR experiences, \rr{TactDeform advances tactile rendering from uniform contact detection toward feature-sensitive spatial feedback.} We provide an open-source implementation to facilitate adoption and enable future research building on these parametric principles. \rr{TactDeform establishes a foundation} for virtual environments that provide the rich tactile understanding \rr{essential for realistic} object exploration and manipulation.

\begin{acks}
This work was supported by the Australian Government’s National Health and Medical Research Council (NHMRC) under the 2022 Ideas Grant (Grant No. 2021183). Dr. Withana is a recipient of an Australian Research Council Discovery Early Career Researcher Award (DECRA) - DE200100479 funded by the Australian Government. We extend our gratitude to all user study participants for their valuable time and contributions. Additionally, we appreciate the members of the AID-LAB for their insightful feedback and support.
\end{acks}

% \balance
\bibliographystyle{ACM-Reference-Format}
\bibliography{references}

% \appendix

\end{document}